\documentclass[aps,prl,reprint,amsmath,amssymb,notitlepage,citeautoscript,superscriptaddress]{revtex4-1}

\usepackage{bm,mathrsfs,dcolumn,graphicx,color}
\usepackage{amsmath}
\usepackage{graphicx}
\usepackage[T1]{fontenc}
\usepackage{hyphenat}
\usepackage{setspace}

\usepackage[normalem]{ulem}
\definecolor{darkerblue}{rgb}{0,0,0.75}
\definecolor{darkerred}{rgb}{0.8,0,0}
\usepackage[colorlinks=true,citecolor=darkerblue,linkcolor=darkerred]{hyperref}

\newcommand{\supp}{See Supplemental Material at [url], which includes Refs.~[72]--[78], for additional experimental details and theoretical analysis.}

\definecolor{ablue}{rgb}{0.1,0.3,0.65}

\begin{document}

\title{Exciton diffusion and halo effects in monolayer semiconductors}

\author{Marvin Kulig}
\author{Jonas Zipfel}
\author{Philipp Nagler}
\author{Sofia Blanter}
\author{Christian Sch\"uller}
\author{Tobias Korn}
\author{Nicola Paradiso}
\affiliation{Department of Physics, University of Regensburg, Regensburg D-93053, Germany}
\author{Mikhail M. Glazov}
\affiliation{Ioffe Institute, Saint Petersburg, Russian Federation}
\author{Alexey Chernikov\footnote{alexey.chernikov@ur.de}}
\affiliation{Department of Physics, University of Regensburg, Regensburg D-93053, Germany}

\begin{abstract}

We directly monitor exciton propagation in freestanding and SiO$_2$-supported WS$_2$ monolayers through spatially- and time-resolved micro-photoluminescence under ambient conditions.
We find a highly nonlinear behavior with characteristic, qualitative changes in the spatial profiles of the exciton emission and an effective diffusion coefficient increasing from 0.3 to more than 30\,cm$^2$/s, depending on the injected exciton density.
Solving the diffusion equation while accounting for Auger recombination allows us to identify and quantitatively understand the main origin of the increase in the observed diffusion coefficient.
At elevated excitation densities, the initial Gaussian distribution of the excitons evolves into long-lived halo shapes with $\mu$m-scale diameter, indicating additional memory effects in the exciton dynamics.
\end{abstract}
\maketitle

Coulomb-bound electron-hole pairs, or excitons, have been in the focus of the solid-state research for many decades~\cite{Frenkel1931, gross:exciton:eng}.
They are of paramount importance for the fundamental understanding of interacting charge carriers in semiconductors~{\cite{ivchenko05a,Klingshirn2007,Haug2009}.
A number of increasingly advanced concepts, including exciton-polaritons~\cite{Weisbuch1992}, Rydberg excitons~\cite{Kazimierczuk2014}, entangled photons from biexcitons~\cite{shields06,Dousse2010}, dropletlike states~\cite{jeffries1983electron,Almand-Hunter2014}, exciton spin currents~\cite{PhysRevLett.110.246403}, and high-temperature Bose-Einstein condensates~\cite{moskalenko,Fogler2014a} among others highlight a vibrant field of ongoing research. 
Recently, excitons in single layers of semiconducting transition-metal dichalcogenides (TMDCs)~\cite{Novoselov2005, Mak2010, Splendiani2010} were found to combine several key traits relevant for both fundamental many-body physics and future technology~\cite{Yu2015,Xiao2017,Wang2018}.
They are unusually stable with binding energies on the order of 0.5\,eV due to strong quantum confinement and weak dielectric screening\,\cite{Rytova1967,Keldysh1979,Cudazzo2011,Berkelbach2013,Qiu2013}, couple efficiently to light\,\cite{Zhang2014a,Poellmann2015}, and can be individually addressed by their valley and spin configuration\,\cite{Xu2014}.
These properties and related phenomena have been extensively studied for the last few years.

In this context, it is interesting to consider that excitons in TMDCs are also \textit{free to move in two dimensions} in close analogy to quantum well systems~\cite{Smith1988,Hillmer1988,Steininger1996}.
This has major implications, including the potential to deliberately manipulate exciton currents as well as to address the interplay between propagation and many-particle interactions.
Moreover, to realize some of the more intriguing concepts mentioned above using excitons at room temperature, understanding and controlling their spatial degree of freedom is a crucial component.
However, \textit{exciton transport} in TMDC monolayers received only little attention beyond initial reports of individual diffusion coefficients~\,\cite{Kumar2014,Mouri2014,Yuan2017} and a recent study emphasizing impurity- and phonon-scattering~\cite{Kato2016}. 
Consequently, there is a strong motivation to systematically explore the physics of exciton propagation in two-dimensional TMDCs.

Here, we address this topic by directly monitoring the spatial behavior of excitons in freestanding and supported single layers (1L) of WS$_2$, a prototypical TMDC, through spatially- and time-resolved photoluminescence (PL) microscopy.
We find a highly nonlinear propagation with the effective diffusion coefficient varying over as much as two orders of magnitude depending on the injected exciton density, accompanied by characteristic changes of the emission profiles.
We identify the main source of this nonlinearity and show that it can be quantitatively understood by including Auger processes into the diffusion equation.
Interestingly, additional memory effects are found to play an important role, as evidenced by the observation of halolike shapes with $\mu$m-scale diameters in the emission~\footnote{We note that during the review process of our manuscript halo-shaped features were reported in the preprint by T.B. Arp et al. (arXiv:1711.06917) in the photocurrent response of MoTe$_2$ and interpreted as an evidence of a condensate}.

\begin{figure}[t]
	\centering
		\includegraphics[width=8.2 cm]{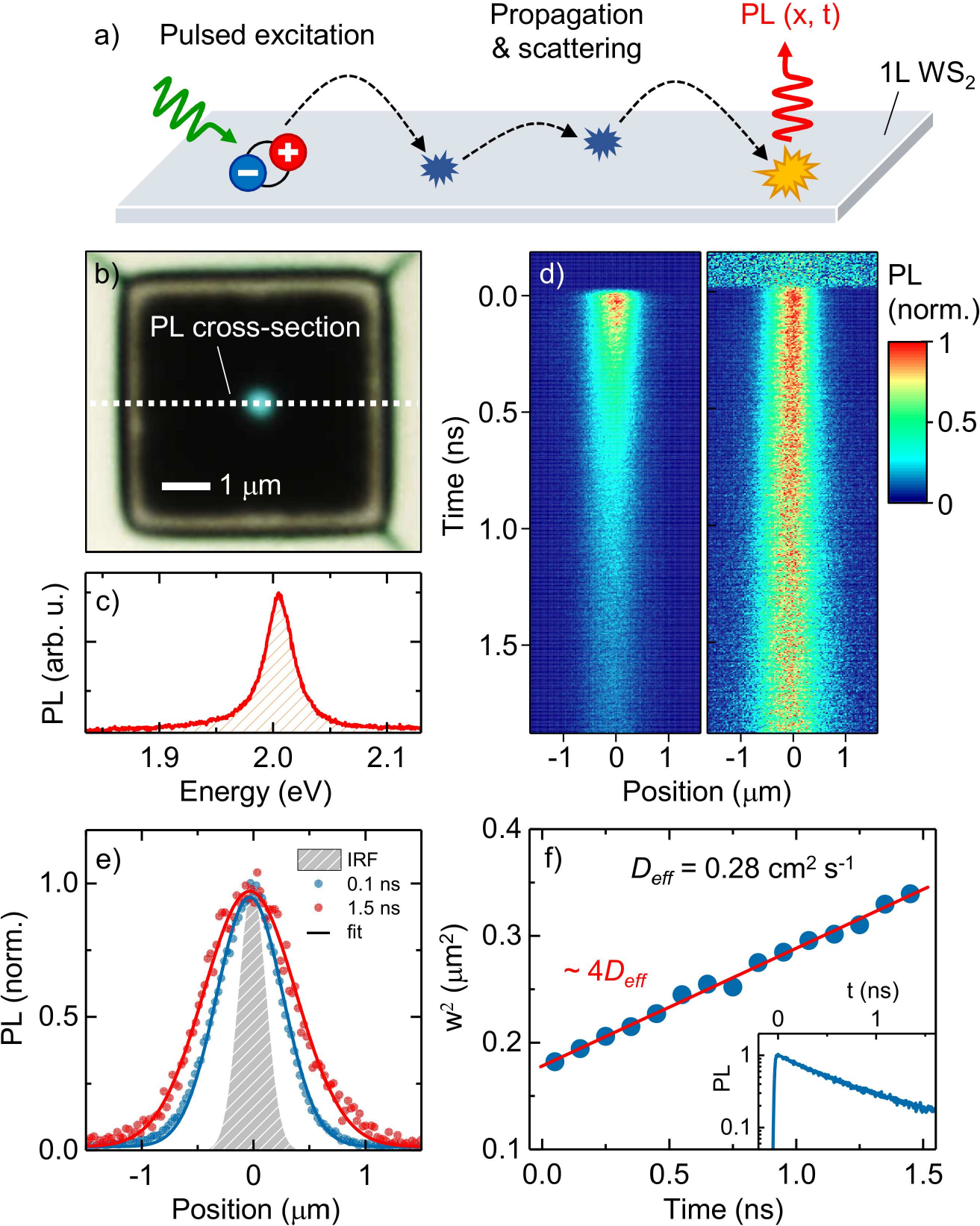}
		\caption{(a) Schematic illustration of the exciton propagation. 
		(b) Optical micrograph of the freestanding 1L\,WS$_2$ sample and the excitation laser spot.
		The dotted line indicates the cross-section of the PL signal imaged onto the detector.
		(c) Luminescence spectrum of the freestanding sample.
		(d) Streak camera image of the PL intensity cross-section	of the freestanding sample; as-measured and normalized data are shown in the left and right panels, respectively.
		(e) Exemplary luminescence profiles together with Gaussian fits. 
		(f) Extracted squared width of the PL as function of time after the excitation. 
		The inset shows the time-dependent intensity.		
		}
\label{fig1}
\end{figure}

The samples under study are mechanically exfoliated from bulk crystals and subsequently transferred using the technique from Ref.~\onlinecite{Castellanos-Gomez2014a} either onto SiO$_2$/Si substrates or 5x5\,$\mu$m$^2$ holes cut into thin SiN membranes, providing supported and freestanding samples, respectively.
A 100\,fs\,-\,pulsed Ti:sapphire laser with a repetition rate of 80\,MHz is used as an excitation source.
The laser is tuned to a photon energy of 2.43\,eV by second-harmonic generation and focused to a spot with a full-width-at-half-maximum of about 0.5\,$\mu$m, creating electron-hole pairs in WS$_2$.
The excitons then form on ultra-short timescales~\cite{Ceballos2016,Steinleitner2017} and are distributed among optically bright and dark states~\cite{Withers2015,Arora2015,Zhang2015a,Wang2015c,Zhang2017,Wang2018}.  
As schematically illustrated in Fig.\,\ref{fig1}\,(a), the excitons propagate, scatter, and a fraction of them subsequently recombines radiatively at a finite distance from the initial injection position.
The latter allows us to optically trace the dynamics of the exciton distribution.

In the experiment, the resulting emission is imaged along the cross-section of the excitation profile as indicated in Fig.\,\ref{fig1}\,(b) and subsequently deflected by either a mirror or a grating to monitor either spatially- or spectrally-resolved signals.
A typical PL spectrum is presented in Fig.\,\ref{fig1}\,(c), exhibiting a single resonance from neutral excitons in WS$_2$~\cite{Zhao2013}.
The luminescence is detected using a streak camera operating in single-photon counting mode.
All experiments are conducted at room temperature under ambient conditions.
Further details are given in the Supplemental Material~\cite{Supplemental}.

A typical spatially-resolved streak camera image of the exciton PL from freestanding 1L\,WS$_2$ is presented in the left panel of Fig.\,\ref{fig1}\,(d) for an average excitation density of 14\,nJ\,cm$^{-2}$.
The signal decays with time on a nanosecond-scale and broadens along the spatial coordinate.
The broadening is further illustrated in the right panel of Fig.\,\ref{fig1}\,(d), where the data is normalized to the intensity maximum at each time step.
Time-dependent spatial profiles are extracted from the image by integrating over intervals of 0.1\,ns on the time-axis.
Exemplary data are presented in Fig.\,\ref{fig1}\,(e) together with the instrument response.
For quantitative analysis, the luminescence intensity $I_{PL}(x,t)$ is fitted using a Gaussian function $\exp{[-x^2/w^2(t)]}$ at each time-step $t$.
The squared width $w^2(t)$, plotted in Fig.\,\ref{fig1}\,(f), increases linearly with time as the excitons propagate and recombine; the PL intensity is plotted in the inset.
The \textit{effective diffusion coefficient} $D_{\rm eff}$ is extracted from the slope according to $w^{2}(t)\,=\,w_0^2+4D_{\rm eff}t$.

The procedure is then repeated while tuning the energy density of the excitation pulse between 1~nJ\,cm$^{-2}$ and 1~$\mu$J\,cm$^{-2}$.
Assuming 9\% absorption at the laser energy, 1~nJ\,cm$^{-2}$ corresponds to an exciton density of $2.3\times10^{8}$\,cm$^{-2}$, constituting the low-density limit of creating less than one electron-hole pair on average per pulse.
The resulting effective diffusion coefficients are summarized in Fig.\,\ref{fig2}\,(a); individual data sets at selected densities are presented in Fig.\,\ref{fig2}\,(b).

In both freestanding and supported samples, the low-density values of $D_{\rm eff}$ converge around 0.3\,cm$^{2}$/s, corresponding to an effective exciton mobility of $eD_{\rm eff}/k_BT$\,=\,12\,cm$^{2}$/Vs and a mean diffusion length of $2\sqrt{D_{\rm eff}\tau}$\,=\,360\,nm, using a recombination time $\tau$\,=\,1.1\,ns for the freestanding sample. 
This result is below rough estimates from a basic kinetic model, yielding diffusion coefficients on the order of $k_BT(\tau_{\rm s}/M_X)$\,=\,2\,cm$^{2}$/s, when scattering times $\tau_{\rm s}$ around 30\,fs~\cite{Selig2016} for the \textit{bright} exciton in WS$_2$ with a total mass of $M_X$\,=\,0.67\,$m_0$~\cite{Kormanyos2014} are assumed.
We note, however, that the applicability of the above estimation is limited for high scattering rates and comparatively low thermal velocities of the excitons in WS$_2$ yielding a mean free path comparable to the exciton radius and the de Broglie wavelength.
Also, the majority of the excitons are optically dark with scattering rates not readily accessible.
In the experiment, the values for $D_{\rm eff}$ close to 0.3\,cm$^{2}$/s are consistently obtained in the low-density limit across the studied samples, independent of the presence of the SiO$_2$/Si substrate (see also Supplemental Material~\cite{Supplemental}).
For comparison, typical room-temperature exciton diffusion coefficients in molecular solids are orders of magnitude smaller\,\,\cite{Akselrod2014,Tamai2015}, whereas the values for quantum wells can be much higher\,\,\cite{Smith1988,Hillmer1988}.
The measured diffusion lengths in 1L~WS$_2$ are roughly on the scale of values reported for carbon nanotubes~\cite{Moritsubo2010} and on the lower end of organolead halide perovskites~\cite{Stranks2013}.  

\begin{figure}[t]
	\centering
		\includegraphics[width=8.4 cm]{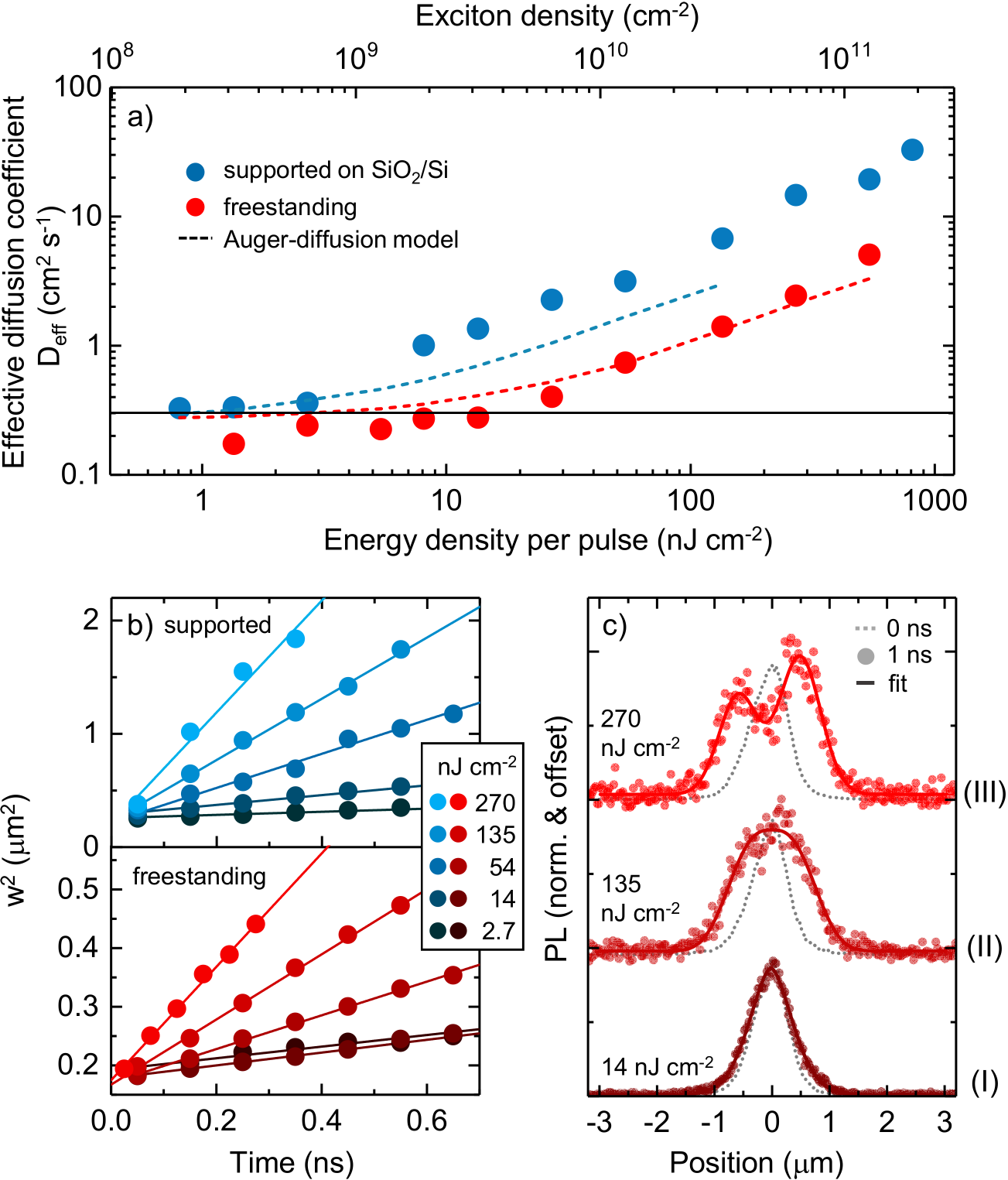}
		\caption{(a) Effective diffusion coefficients in freestanding and supported 1L\,WS$_2$ samples together with the simulation results from Auger-diffusion model. 
		(b) Squared widths of the spatially-resolved PL as function of time including linear fits.
		(c) Spatial PL profiles of the freestanding sample after 1\,ns, typical for the three density regimes.
		}
\label{fig2}
\end{figure}

Remarkably, at elevated exciton densities, we observe a large increase of the effective diffusion coefficient over two orders of magnitude, reaching up to 30\,cm$^{2}$/s (see Fig.\,\ref{fig2}\,(a)).
Previously reported individual diffusion coefficients for supported TMDC monolayers fall roughly into the middle of this range~\,\cite{Kumar2014,Mouri2014,Yuan2017}. 
Interestingly, the increase of $D_{\rm eff}$ is further accompanied by qualitative changes of the emission shape, as illustrated in Fig.\,\ref{fig2}\,(c).
While the spatial profiles of the PL cross-section remain Gaussian in the linear diffusion regime, they acquire a more pronounced flat-top character, resembling super-Gaussian peak functions $\exp[-\left|x/w(t)\right|^p]$ with $p>2$, and finally evolve into a double-peak at later times at higher densities.
For the data presented in Fig.\,\ref{fig2}\,(a), the effective diffusion coefficient is extracted in the initial time range after the excitation where fitting by a single peak function is largely applicable.
Here, we emphasize that the studied density range corresponds to rather moderate excitation conditions, far below carrier concentrations where pronounced many-particle renormalization effects are expected~\cite{Wang2018}.
This is further supported by the lack of measurable energy shifts and broadening in spectrally-resolved PL, also indicating negligible heating (see Supplemental Material~\cite{Supplemental}). 
We note that similar behavior is observed in WSe$_2$ monolayers as presented in the Supplemental Material~\cite{Supplemental}.

\begin{figure}[t]
	\centering
		\includegraphics[width=8.4 cm]{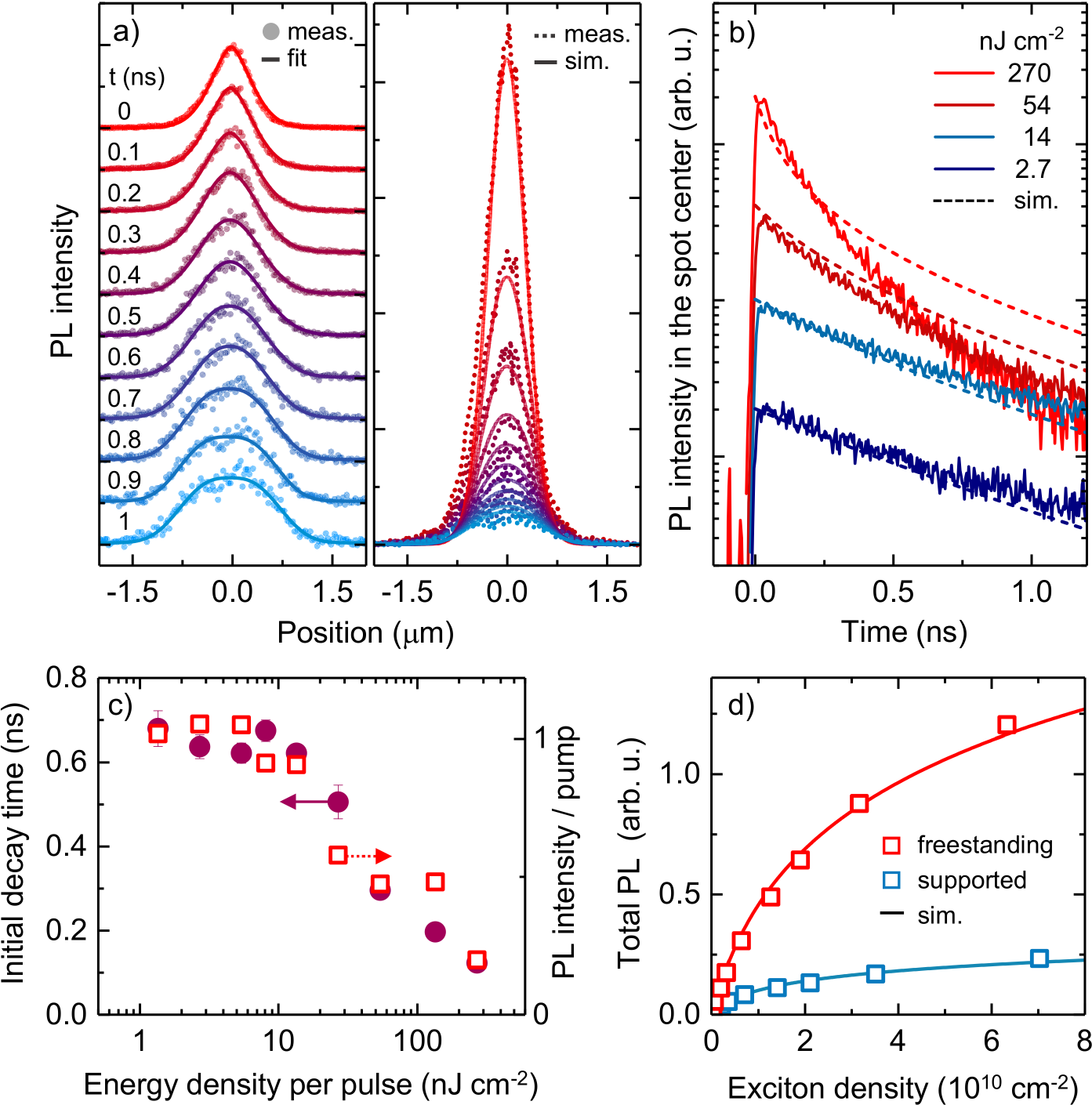}
		\caption{(a) PL profiles of the freestanding sample for an excitation density of 135\,nJ\,cm$^{-2}$.
		The data is shown normalized and offset with super-Gaussian fit curves (left panel) and as-measured, together with the simulation results (right panel).
		(b) PL transients from the center of the excitation spot of the freestanding sample and the simulation results. 
		(c) Initial decay time of the luminescence and the relative PL yield within the cross-section.
		(d) Total time-integrated PL intensity.
		}
\label{fig3}
\end{figure}

In the following, we focus on the nonlinear regime at elevated densities.
Taking the freestanding data set as an example, a series of PL profiles at different times after the excitation with the intermediate density of 135\,nJ\,cm$^{-2}$ are presented in Fig.\,\ref{fig3}\,(a).
As-measured data is shown in the right panel, the profiles on the left are normalized and offset, including fits by super-Gaussians illustrating the continuous evolution of a flat-top.
In addition, we observe a decrease in the decay time and relative luminescence intensity with increasing density beyond the linear regime, as shown in Fig.\,\ref{fig3}\,(b).
The initial decay constant after the excitation and the relative PL yield within the detected cross-section are presented in Fig.\,\ref{fig3}\,(c).
Correspondingly, the total PL intensity, obtained by imaging the emission onto a charge-coupled-device camera and plotted in Fig.\,\ref{fig3}\,(d), saturates with increasing density. 

Simultaneous decrease of the relative PL yield and recombination time are hallmarks of non-radiative Auger recombination, often labeled as exciton-exciton annihilation and observed in various TMDC monolayers~\cite{Kumar2014,Mouri2014,Sun2014,Yu2016,Yuan2017,Hoshi2017,Manca:2017aa}.
When two excitons interact, one of them can recombine, transferring the energy to the other and exciting it to a state at higher energies.
The probability of this process increases with the exciton density $n$ and the recombination rate is usually presented as a bimolecular decay $R_An^2$ with the Auger coefficient $R_A$.
Hence, as the excitons recombine faster at elevated densities in the middle of the spot, the profile should become increasingly flat. 
With time, this should lead to an effective additional broadening of the exciton distribution and result in an apparent increase of the diffusion coefficient, as also observed in CuO$_2$ bulk crystals\,\cite{Warren:2000} (in contrast to the repulsion of indirect excitons in GaAs double quantum-wells\,\cite{Voros2005}).

To show that the interplay between exciton-exciton interactions and propagation can largely account for the experimental observations, we introduce Auger recombination into the diffusion equation:
\begin{equation}
\label{diffusion}
\frac{\partial n}{\partial t} = D\Delta n - \frac{n}{\tau} - R_A n^2,
\end{equation}
where $D$ and $\tau$ are the low-density diffusion coefficient and the recombination time, respectively, and $\Delta$ is the Laplace operator.
We fix the parameters to the values: $D$\,=\,$D_{\rm eff}(n\rightarrow 0)$\,=\,0.3\,cm$^{2}$/s; time constants $\tau$\,=\,1.1\,ns (freestanding) and $\tau$\,=\,0.7\,ns (supported) which are self-consistently extracted from the exponential PL decay in the linear regime, taking into account the additional decay channel from diffusion.
The coefficients $R_A$\,=\,0.14\,cm$^{2}$/s (freestanding) and $R_A$\,=\,0.5\,cm$^{2}$/s (supported) are chosen within the measured limits from both the relative increase of the recombination rate and the saturation of the total PL (also see Supplemental Material~\cite{Supplemental}).
They are consistent with previous reports~\cite{Kumar2014,Mouri2014,Sun2014,Yu2016,Yuan2017,Hoshi2017}. 
The biexciton formation discussed in Ref.~\cite{Wolfe2014} can be also included in Eq.~\eqref{diffusion} by renormalizing $R_A$.
 
With all parameters fixed, Eq.~(1) is numerically solved, using a Gaussian profile of the size of the PL spot immediately after the excitation ($w_0=0.4$\,$\mu m$) and the injected exciton densities from the experiment as initial conditions.
The results for the profiles presented in Fig.\,\,\ref{fig3}\,(a), transients in Fig.\,\ref{fig3}\,(b), and the total PL intensity in Fig.\,\ref{fig3}\,(d) are plotted alongside experimental data, showing reasonable agreement.
More importantly, the observed increase of the effective diffusion coefficient presented in Fig.\,\,\ref{fig2}\,(a) is reproduced by the simulation, as evaluated during the first 100\,ps after the excitation.
Both the relative shift of the onset and the density-dependent slope of $D_{\rm eff}$ are essentially captured by the model with some overestimation of the onset for the supported sample.
The Auger-diffusion equation also yields approximate analytic expressions for a number of relevant observables such as the time-dependent emission, total PL intensity, and the effective diffusion coefficient (see Supplemental Material~\cite{Supplemental}).
The latter has the form $D_{\rm eff}\approx D+R_A n_0 w_0^2/16$ with $n_0$ and $w_0$ being the initial peak density and the width of the exciton distribution, respectively.

\begin{figure}[t]
	\centering
		\includegraphics[width=8.4 cm]{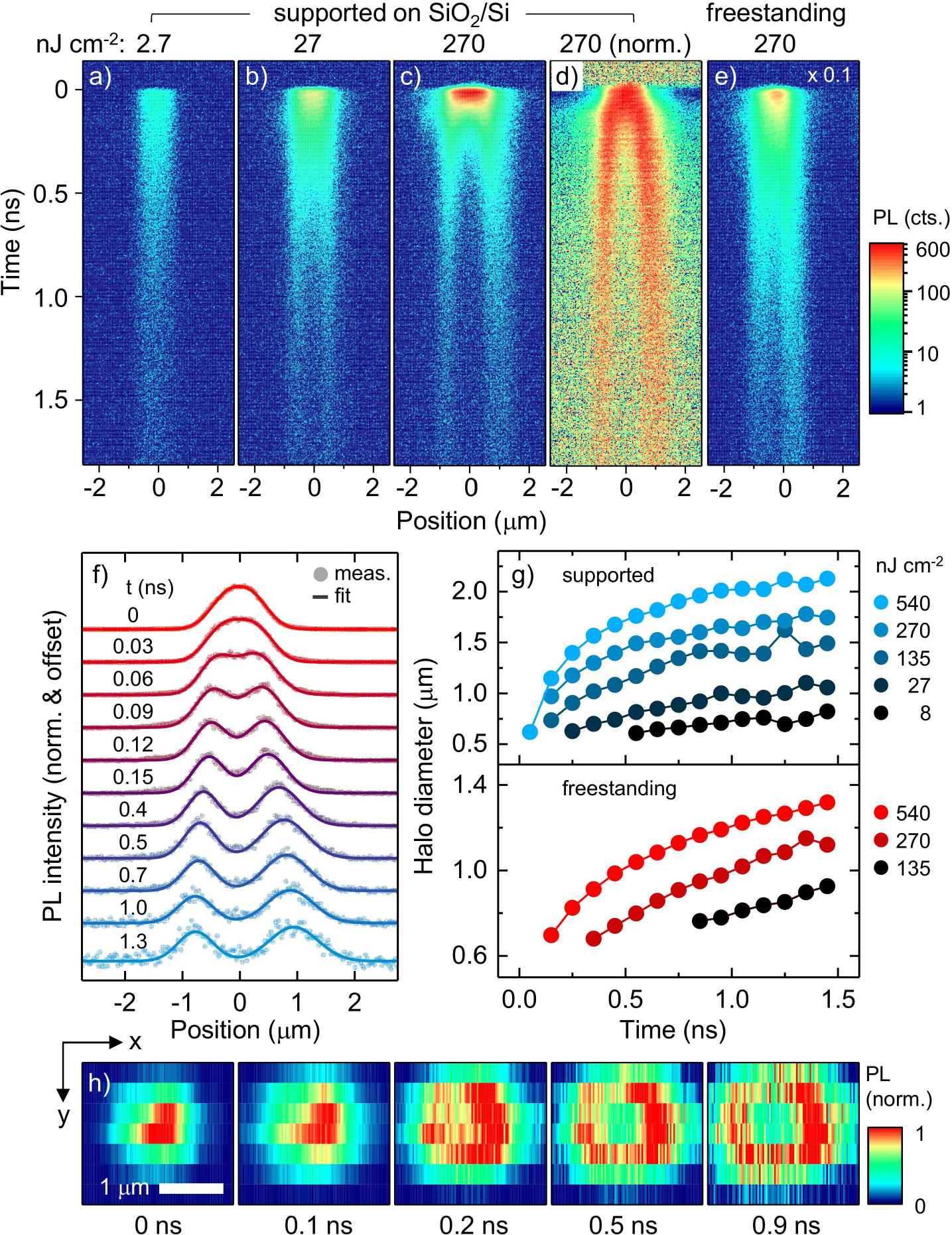}
		\caption{(a)\,-\,(c) As-measured streak camera images of the emission from the supported 1L\,WS$_2$ sample. 
		(d) Same as (c), normalized to the intensity maximum at each time step.
		(e) As-measured PL image of the freestanding sample in the exciton halo regime.
		(f) Luminescence profiles at different times after the excitation of the supported sample from Fig.\,(d), shown together with double-Gaussian fits.
		(g) Diameter of the exciton halo as function of time for supported (top panel) and freestanding (bottom panel) samples.
		(h) Two-dimensional PL images from the supported sample (270\,nJ\,cm$^{-2}$) at different times after the excitation.
		}
\label{fig4}
\end{figure}

While the basic level of theory allows us to identify the main origin of the increasing effective diffusion, notable deviations remain at elevated densities and at longer times after the excitation, see, e.g., the tail of the transients in Fig.\,\,\ref{fig3}\,(b).
A particularly peculiar observation in this regime is the evolution of the exciton emission into halolike shapes, as shown in Fig.\,\,\ref{fig2}\,(c) and in the streak camera images in Fig.\,\,\ref{fig4}\,(a)\,-\,(e).
Individual PL profiles together with double-Gaussian fits are presented in Fig\,\,\ref{fig4}\,(f), highlighting the gradual change.
The double-peak structure in the PL cross-section is observed for all studied samples, regardless of the angle and position, and independent of the presence of the substrate.
The halos remain stable and slowly expand with diameters on the order of 1\,$\mu$m, as illustrated in Figs.~\ref{fig4}\,(g) and (h), showing the halo diameter and two-dimensional time-resolved PL images, respectively. 
We also note that in contrast to valley-related phenomena at cryogenic temperatures~\cite{Rivera2016, Onga2017}, no pronounced polarization dependence is detected (see Supplemental Material~\cite{Supplemental}).
Finally, no lasting effects are observed after decreasing the excitation density back to low values.  

It is instructive to consider that introducing an additional $n$-dependent nonlinearity in $R_A$, $\tau$, or $D$} into the diffusion equation~\eqref{diffusion} does not lead to the evolution of a halo.
The diffusive current is always driven by the density gradient and is directed to smoothen it. 
Hence, as the profile becomes flat, resembling the $t$\,=\,0.06\,ns trace in Fig.\,\,\ref{fig4}\,(f), there is no apparent reason for the density in the center to decrease below a value in close spatial proximity.
We also note that an instability towards non-monotonic profile formation may occur in multi-component nonlinear systems~\cite{Turing:1952aa}, as, e.g., discussed for the interplay of free carriers and excitons in GaAs-based double quantum wells~\cite{Butov2002,Snoke2002,PhysRevLett.94.176404,Rapaport2004, Ivanov2006, Stern2008} at cryogenic temperatures.
In the present case, however, such processes are not likely to contribute for strongly-bound excitons at room temperature in the density regimes far below the Mott transition, with large binding energies and fast formation.

Instead, a memory component seems to be required, so that the exciton behavior becomes dictated by previous events.
A good candidate is the Auger process itself, since the remaining excitons gain large amounts of energy and subsequently experience a number of scattering processes resulting, e.g., in a higher exciton temperature or additional recombination.
Indeed, the appearance of the halo follows the injected exciton density and depends on the sample geometry similar to Auger recombination.
A possible scenario involving overheated excitons is discussed in the Supplemental Material~\cite{Supplemental}; for effects related to ballistic phonons, see Refs.~\cite{Zinovev1983,Bulatov1992,Tikhodeev1998}.
A theoretical full many-body treatment of excitonic scattering and carrier relaxation, however, would be required for an adequate microscopic description of these findings.

In conclusion, we have systematically studied exciton propagation in atomically thin WS$_2$ monolayers.
We find a strong increase in the effective diffusion coefficient over two orders of magnitude due to the interplay of exciton interactions and diffusion.
Our results provide direct access to the inherent diffusion of the excitons and establish a basis for the interpretation of exciton transport experiments in this field.
The presented Auger-diffusion model is easy to implement and captures the main characteristics of the studied behavior in the intermediate density regime.
In addition, while the appearance of long-lived exciton halos in 2D TMDCs is highly intriguing by itself, the ability to deliberately create $\mu$m-sized ring-shaped emitters in ultra-thin materials should be also interesting for photonics and polaritonics, stimulating further research.


%

\section{Acknowledgments}

We thank Archana Raja and Tony F. Heinz for helpful discussions and Thomas Franzl (Hamamatsu Photonics) for technical advice.
Financial support by the DFG via Emmy Noether Grant CH 1672/1-1, Collaborative Research Center SFB 1277 (B05, B08), and KO3612/3-1 project is gratefully acknowledged. 
M.M.G. is grateful to RFBR grant 16-02-00375 and Russian President grant MD-1555.2017.2 for partial support.

M.K. and J.Z. contributed equally to this work.

\end{document}